\documentclass[preprints,article,accept,moreauthors,pdftex]{Definitions/mdpi} 
\UseRawInputEncoding 
\firstpage{1} 
\pubvolume{xx}
\issuenum{1}
\articlenumber{5}
\pubyear{2019}
\copyrightyear{2019}
\history{Received: date; Accepted: date; Published: date}

\pdfoutput=1

\Title{The role of small scale experiments in the direct detection of dark matter}

\Author{{Susana }Cebri\'an $^{1,}$$^{2}$\orcidA{}}

\AuthorNames{Susana Cebri\'an}

\address{%
$^{1}$ \quad Centro de Astropart\'iculas y F\'isica de Altas Energ\'ias (CAPA), Universidad de Zaragoza, C/ Pedro Cerbuna 12, 50009 Zaragoza, Spain; scebrian@unizar.es \\ 
$^{2}$ \quad Laboratorio Subterr\'aneo de Canfranc, Paseo de los Ayerbe s/n, 22880 Canfranc Estaci\'on, Huesca, Spain}

\abstract{In the direct detection of the galactic dark matter, experiments using cryogenic solid-state detectors or noble liquids play for years a very relevant role, with increasing target mass and more and more complex detection systems. But smaller projects, based on very sensitive, advanced detectors following new technologies, could help in the exploration of the different proposed dark matter scenarios too. There are experiments focused on the observation of distinctive signatures of dark matter, like an annual modulation of the interaction rates or the directionality of the signal; other ones are intended to specifically investigate low mass dark matter candidates or particular interactions. For this kind of dark matter experiments at small scale, the physics case will be discussed and selected projects will be described, summarizing the basics of their detection methods and presenting their present status, recent results and prospects.}


\begin{document}


\section{Introduction}

According to a great deal of cosmological and astrophysical observations {at different scales, there is a need of both dark energy and dark matter (DM) in the Universe; although the dark matter particles could explain an important fraction of the its energy-mass budget, its nature is unknown} \cite{RevModPhys.90.045002,pdg2020}. A plethora of candidates have been proposed to form the galactic DM, spanning over tens of orders of magnitude both in mass and interaction cross sections; axions and Weakly Interacting Massive Particles (WIMPs) are ones of the preferred candidates. Different approaches are followed in the investigation of the DM particles, based on the direct detection \cite{natphysdir}, indirect searches after annihilation \cite{natphysindir} or production at accelerators \cite{natphysacc}.

The direct detection of DM was focused for years on the elastic scattering off target nuclei of WIMPs, for spin-independent (SI) {and} spin-dependent (SD) interactions, {giving rise to nuclear recoils (NR) as detectable signals}; but inelastic scattering {and also scattering with electrons in the detector medium producing electron recoils (ER) are being considered too}, as it will be shown later. {Effective field theories (EFTs) including a minimal set of interactions and particles are being used in some cases as the frame to derive experimental results considering a number of operators describing the various possible interactions in a general way} \cite{Fitzpatrick_2013}. {Assuming a particular type of interaction, the expected counting rates can be predicted considering certain astrophysical parameters (like the local density, the velocity distribution and the escape velocity of the DM particles) and properties of the DM particle} \cite{drukier}; the direct detection formalism is summarized, for example, in the Dark Matter section of \cite{pdg2020} {and detailed for elastic NR in} \cite{LEWIN199687}.

This {direct} detection mechanism is really challenging as the produced signal is rare, with very low energy and has a continuum energy spectrum (which decays almost exponentially) making it indistinguishable from backgrounds \cite{Schumann:2019eaa}. The DM signal is searched in the detectors at the keV-scale and below and the expected interaction rates are typically lower than 1 event per kg and year. Therefore, the two basic requirements for this type of experiments are:
\begin{itemize}
    \item Very low energy threshold.
    \item Ultra low background conditions. As for other rare event searches, operation deep underground to suppress cosmic rays and the use of passive and active shieldings against the environmental radiation from radioactivity and neutrons are mandatory \cite{heusser,formaggio}. Careful control of the material intrinsic radiopurity {(in bulk or on surface)} and of the cosmogenic activation of components producing long-lived radionuclides is also a must \cite{cebriancosmogenic,universecosmogenic}. {Appropriate materials must be selected based on different types of radioassays and purification techniques are often applied.} The implementation of specific background rejection techniques, {following for instance ER/NR discrimination by measuring different observables (heat, light and charge), Pulse Shape Discrimination (PSD) or volume fiducialisation if spatial information of events is available,} has allowed to reduce the background levels in some experiments event down to 10$^{-4}$~c/keV/kg/d, leaving neutrinos (from the Sun, supernovae or the atmosphere) as an irreducible background \cite{PhysRevD.89.023524}.
\end{itemize}
{The accumulation of a large quantity of target material and very long data taking periods expanding for months or years are also required, together with the stability of the environmental conditions and, in general, of the response of the detector along all that period.} 
Additionally, the detection of distinctive signatures, such as an annual modulation in the rates of interaction or the signal directionality, would be priceless to attribute a DM origin to a potential observation. 

Different types of experiments based on different, complementary technologies are underway, using for instance bolometers, liquid Argon or Xenon detectors, scintillators, semiconductors, gas detectors or bubble chambers \cite{Schumann:2019eaa,Undagoitia_2015}. 
Experiments based on bolometers and on noble liquid detectors with continuously increasing target mass have released in the last years the strongest limits on DM interaction cross sections following the direct detection approach. But initiatives for smaller projects based on different technologies are being now strongly supported by the community \cite{whitepaper}, as they can help to investigate well-motivated candidates, particularly for masses below the GeV/c$^2$ scale. The aim of this work is to review the goals and status of these {other} direct detection projects. The structure of the paper is the following. In the rest of this section, a brief summary of the latest results and prospects of experiments using liquid Argon and Xenon detectors and bolometers will be given for the sake of completeness (as they are not addressed more in the review). Running experiments and projects intended to identify distinctive signatures of the DM interaction will be presented in Sec.~\ref{secam}, for the annual modulation in the expected rates, and in Sec.~\ref{secdir}, for the signal directionality. Those intended to specifically explore low mass DM using purely ionization detectors, like gas chambers or semiconductors, will be discussed in Sec.~\ref{seclm} while  Sec.~\ref{secsd} will be devoted to experiments focused on SD interactions. Finally, summary and outlook will be given in Sec.~\ref{secsum}.

\subsection{Liquid Ar and Xe detectors}
In noble gases, like Argon and Xenon, both scintillation and ionization are produced. They are being used in liquid as massive and compact DM targets, which are producing dominant results for high mass WIMPs, from a few GeV/c$^2$ to TeV/c$^2$. A dual-phase liquid noble gas detector is a Time Projection Chamber (TPC) which can register by means of adequate photosensors both the primary (S1) and the secondary (S2) scintillation from drifted electrons; then, the ratio S1/S2 can be used to discriminate {ER from NR}, with a threshold around 1~keV$_{ee}$\footnote{Electron equivalent energy.}. Additionally, a 3D event reconstruction is possible with spatial resolution at the order of mm (using the coordinate deduced from the drift time). Targets at the level of tens, hundreds and even tons of kg have been already operated; the chamber is at the heart of the detection systems, accompanied by huge radiation shields, veto detectors, cryogenic and gas purification systems, elaborated data acquisition equipment and other ancillary systems.

Three large dual-phase Xenon experiments have presented in the last years the leading constraints on SI DM-nucleus interactions for masses higher than  6~GeV/c$^2$: the XENON1T experiment at Laboratori Nazionali del Gran Sasso (LNGS),  in Italy \cite{Aprile:2018dbl}, LUX, already finished, at the Sanford Underground Research Facility (SURF), in US \cite{Akerib:2016vxi}; and the PANDAX-II experiment operated at Jinping Underground Laboratory, in China \cite{Cui:2017nnn,pandaxiifull}. XMASS-I operates a single-phase liquid Xenon detector at the Kamioka Observatory in Japan \cite{201945}. Using only the S2 signal, results for DM-electron scattering have also been obtained by XENON1T with a threshold of 0.4~keV$_{ee}$ \cite{Aprile:2019xxb} and by PANDAX-II \cite{pandaxiie}. Specific investigation of light DM, being enhanced by the Migdal effect \cite{Ibe:2017yqa} or Bremsstrahlung when searching for ER, has been carried out by XENON1T \cite{Aprile:2019jmx} and by LUX \cite{Akerib:2018hck}. In addition, a wide range of different DM models has been taken into account by the collaborations (effective interactions, inelastic scattering, light mediators, bosonic Super-WIMPs, dark photons $\dots$); axions have been considered when looking for the origin of the ER excess reported by XENON1T \cite{Aprile:2020tmw}. The presence in natural Xenon of $^{129}$Xe and $^{131}$Xe, with non-null spin, makes these detectors sensitive also to SD interactions. Extensions of the three dual-phase experiments, using tonnes of Xenon active mass at the same locations, are now being prepared or commissioned with expected operation along the next years: XENONnT, LUX-ZEPLIN (LZ) and PANDAX-4T. The running of the DARWIN observatory in preparation, with 40~tonnes of Xenon, is foreseen for the end of the decade.

In Argon detectors, a very efficient PSD is possible for ER and NR. DEAP-3600, with a single-phase liquid Ar detector at SNOLAB in Canada, has demonstrated an excellent background rejection capability \cite{Ajaj:2019jk}. DarkSide-50 has operated at LNGS a dual-phase liquid Ar detector, filled with underground Ar (UAr) with strongly suppressed (by $\sim$1400) $^{39}$Ar activity \cite{Agnes:2018ep}; detecting the S2 signal only and achieving a 100~eV$_{ee}$ energy threshold, the search for low mass DM gave the strongest limits from 1.8 to 3.5~GeV/c$^2$ for SI DM-nucleus interaction \cite{Agnes:2018fg} and also results on DM-electron interaction \cite{Agnes:2018ft}. Effective field theory interactions have been also considered in the analysis of these Argon detectors. The Global Argon DM Collaboration (GADMC) has been constituted to cooperate for the procurement of radiopure UAr from the Urania plant in Colorado, US, and the Aria facilities in Sardinia, Itay, as well as for the development of  SiPMs for readout. The DarkSide-20k detector will operate, as the first step, 20~tonnes of active mass of UAr inside a 3.5~m long TPC made with acrylic walls and placed inside an  atmospheric Ar veto at LNGS; a smaller detector optimized for low mass DM investigation with just 1~tonne of purified UAr (DarkSide-LowMass) is also in preparation. In a longer term, the ARGO project, with 300~tonnes of UAr, could be in operation at SNOLAB at the end of the decade; atmospheric neutrinos could be the limiting background.

\subsection{Bolometers}
In solid-state detectors operated at temperatures of tens of mK o even below, phonons (heat) can be registered by the very small increase of temperature induced; the simultaneous detection of scintillation (light) or ionization (charge) signals makes possible the discrimination between NR and ER. No quench affects the measured heat signal in these detectors. The operation temperature requires the use of powerful cryogenic systems to cool down even tens of kg of crystals. Very low energy thresholds below 100~eV$_{nr}$\footnote{Nuclear recoil energy.} have been reached, specially using crystals with very small mass; therefore, this kind of detectors are releasing the stringest results for DM-nucleus scattering in GeV/c$^2$ and sub-GeV/c$^2$ mass regions.

EDELWEISS-III presented very good bounds from 5 to 30 GeV/c$^2$, and also results on Axion-Like Particles (ALPs) \cite{Hehn:2016nll}, operating 24 Germanium bolometers (each one with a mass of 870~g) at the Laboratoire Souterrain de Modane (LSM), in France. EDELWEISS is now working with smalll germanium bolometers (with mass of 33~g), first above ground and then at Modane, having reached a 60~eV energy threshold. Masses down to 45~MeV/c$^2$ when including the Migdal effect have been explored and DM-electron scattering as well as dark photons have been taken into account too \cite{Armengaud:2019kfj,Arnaud:2020svb}.

SuperCDMS has operated Silicon and Germanium bolometers, also at the scale of hundreds of grams each, at the Soudan Underground Laboratory, in US. An energy threshold of 70~eV was obtained, thanks to the NTL (Neganov-Trofimov-Luke) effect (applying high bias voltage to transform ionization to heat); results for DM-nucleus scattering up to 1.5~GeV/c$^2$ and for both SI and SD interactions were obtained from different analyses \cite{Agnese:2017njq,Agnese:2018gze}. Sensitivity to SD scattering is possible thanks to the presence of $^{73}$Ge. More recently, very small Silicon crystals (with masses of 0.93~g and 10.6~g) have been operated on surface, deriving limits on DM-nucleon cross sections and also taking into account DM-electron interactions and dark photons \cite{Amaral:2020ryn,Alkhatib:2020slm}. The beginning of the operation at SNOLAB of SuperCDMS, combining different kinds of Silicon and Germanium detectors working in different modes with a total target mass of $\sim$30~kg, is foreseen for 2021.

Similarly, after using scintillating bolometers of CaWO$_4$ with a mass of 300~g each in CRESST-II \cite{Angloher:2015ewa}, ten CaWO$_4$ crystals with a much smaller mass each (24~g) are being operated at LNGS in CRESST-III. An energy threshold of 30~eV has been obtained, setting the best bounds for SI DM-nucleus interaction up to 160~MeV/c$^2$ and similarly for SD interaction at low mass, particularly for the neutron-only case using also a Li$_2$MoO$_4$ crystal containing $^7$Li and $^{17}$O \cite{Abdelhameed:2019hmk,cresstsd2019}. In the next future, operation of about 100 crystals is planned and the energy threshold could be lowered down to 10~eV.

\section{Annual modulation effect}
\label{secam}

As the Earth moves around the Sun, the relative velocity between the DM particles (within the galactic halo) and the detector (placed in the Earth) follows a cosine dependence; this gives rise to a modulation in time in the DM interaction rate, $S$, which can be expressed for an energy bin $i$ as:
\begin{equation}
S_{i}(t)=S_{0,i}+S_{m,i}\cos(\omega(t-t_{0})),
\end{equation}
\noindent being $S_{0,i}$ the non-modulated component and $S_{m,i}$ the modulation amplitude \cite{freese1988}. Assuming a locally isotropic DM halo, the modulation period $T=2\pi/\omega$ is one year and the phase makes the maximum rate be expected around June, 2$^{nd}$. The effect has well-defined properties: it must be weak (just a few percent variation), noticeable only at low energy and a phase reversal is expected at the very low energy region \cite{drukier,freese1988,freese2013}. As no background component is known to show all these features, its identification is considered a distinctive signature of the DM interaction. It is worth noting that the features of the modulation effect are different for other DM distributions, for instance, in the case of sub-structures like streams and clumps or if considering a DM disc.

Scintillating crystals made of NaI(Tl) and coupled to photomultipliers (PMTs) are perfectly suited for the detection of the DM annual modulation, being robust and affordable detectors; they are capable of running in stable conditions for long times and a large target mass ca be easily accumulated. In any case, to achieve an  ultra-low background (reducing significantly the content of $^{210}$Pb and $^{40}$K in commercial low-background NaI(Tl) detectors) and a low energy threshold, it has been necessary to implement new developments also in this type of conventional detectors.

The DAMA/LIBRA (DArk MAtter/Large sodium Iodide Bulk for RAre processes) experiment has been collecting data at LNGS for more than two decades. NaI(Tl) detectors fabricated by Saint Gobain company (9.7~kg each) are used \cite{Bernabei:2008yi}; nine units were firstly operated and, since 2003, there are 25 modules having a total mass of $\sim$250~kg. All PMTs were replaced in 2011 for the second phase of the experiment, which allowed to reduce from 2 to 1~keV$_{ee}$ the software threshold. In the region of interest, the measured background level goes from 0.5 to 1~c/keV/kg/d~\cite{damaphase2}. The results from the first phase \cite{damaphase1} were corroborated by the ones of the second one \cite{damaphase2,bernabeippnp2020}, reporting the presence of a modulation signal with all the expected properties at 12.9$\sigma$ C.L., following an exposure of 2.46~tonne$\cdot$y collected over 20~years. The modulation amplitude deduced in the 2-6~keV$_{ee}$ window is $S_{m}=(0.0103\pm0.0008)$~c/keV/kg/d; compatible results  were obtained when applying other fitting methods and considering different periods of time, detector units and energy windows from 1 to 6~keV$_{ee}$. After the second phase of DAMA/LIBRA, new model-dependent, corollary analyses have been presented \cite{damacorollary,bernabeippnp2020}; maximum likelihood procedures have been applied to define the allowed regions in the space of parameters of several scenarios, by comparing with the theoretical expectations the derived amplitude for the annual modulation. A third phase of the experiment is now in preparation, implementing hardware improvements with the aim to reduce the threshold below 1~keV$_{ee}$.

Considering the bounds on DM interaction cross sections presented by other direct detection experiments (see for instance the compilation at \cite{pdg2020}), a  tension appears if interpreting the DAMA/LIBRA modulation signal as due to DM, considering standard but also more general halo or interaction models. In experiments using different targets, annual modulation has not been observed neither with xenon \cite{Aprile:2017yea,Akerib:2018zoq,Kobayashi:2018jky} nor with  germanium \cite{cdexmod}. In this situation, to prove or disprove in a model-independent way the DAMA/LIBRA observation using the same target and detection technique would be essential. Various projects have undertaken this task; presently, only COSINE-100 and ANAIS-112 are already taking data.

The ANAIS experiment (Annual modulation with NAI Scintillators) operates at the Canfranc Underground Laboratory (LSC, ``Laboratorio Subterr\'aneo de Canfranc'') in Spain; ANAIS-112 includes nine NaI(Tl) detectors (112.5~kg in total) manufactured by the Alpha Spectra company \cite{anaisperformance}. The DM data taking is running since August 2017, with an excellent light collection at the level of $\sim$15 phe/keV for all detector units \cite{OLIVAN201786}, which allows to work with an energy threshold at 1~keV$_{ee}$. A careful study of the different background components from the data taken with all the detectors, specific measurements and Monte Carlo simulations has been made \cite{anaisjcap,anaisbkg2016,anaisbkg} and discrimination methods for rejecting non-scintillation events are implemented \cite{anaisperformance}; from 1 to 6~keV$_{ee}$, the background level is measured  as 3.6~c/keV/kg/d after efficiency correction for cuts~\cite{anaisbkg}. The sensitivity of ANAIS-112 was assessed from this measured background, confirming the exploration of the 3$\sigma$ region singled out by DAMA/LIBRA from five years of data~\cite{anaissensitivity}. First results from an annual modulation analysis were published considering the first 1.5~years, corresponding to an exposure of 157.55~kg$\cdot$y \cite{anaismod}. Updated results {were} presented for 2~years (with a total exposure of 220.69~kg$\cdot$y) applying the same analysis methods~\cite{anaistaup}. {The analysis of the first 3~years of data (corresponding to 313.95~kg$\cdot$y) has been released, being the modulation amplitudes from best fits obtained $S_{m}=(-0.0034\pm0.0042)$~c/keV/kg/d for the 1-6~keV$_{ee}$ region and $S_{m}=(0.0003\pm0.0037)$~c/keV/kg/d in 2-6~keV$_{ee}$} \cite{anais3y}{; this supports the absence of modulation in the data and the values are incompatible with DAMA/LIBRA result at 3.3 (2.6)~$\sigma$ in the 1-6~keV$_{ee}$ (2-6~keV$_{ee}$) windows, for a sensitivity of 2.5 (2.7)~$\sigma$. Improvements in the background modelling for the fitting of the rates in the region of interest have been implemented in the analysis of the 3~y data. Additionally, several consistency checks and some complementary analyses (a phase-free annual modulation search and the exploration of possible periodic signals at other frequencies) have been performed too at} \cite{anais3y}. {All the obtained results have allowed to fully confirm the expected sensitivity and the data taking is smoothly ongoing.}

The COSINE project was conceived to join efforts between the DM-Ice experiment installed in the South Pole \cite{PhysRevD.95.032006} and the KIMS collaboration. In COSINE-100, eight NaI(Tl) detectors (106~kg in total) built also by the Alpha Spectra company are operated at the Yangyang underground Laboratory in South Korea, being immersed in 2200~l of liquid scintillator (to veto mainly the $^{40}$K emissions from the NaI(Tl) crystals) \cite{cosineperformance,cosinebkg}; the data taking started in September 2016 with an energy threshold of 2~keV$_{ee}$. From the analysis of the first 59.5~days of data collected, COSINE-100 excluded the DAMA/LIBRA signal as due to the SI interaction of WIMPs in a standard halo model \cite{cosinenature}; these data have been also interpreted in other contexts like Inelastic Boosted DM \cite{PhysRevLett.122.131802} and WIMP effective models \cite{cosinejcapeff}. Results for solar axions have also been derived \cite{cosineaxion}. A first analysis of annual modulation from 1.7~years of data has been presented \cite{cosinemod}; since three crystals were excluded (because of a reduced light yield), the total exposure analyzed is 97.79~kg$\cdot$y. The best fit for the modulation amplitude in the 2-6~keV$_{ee}$ gives $S_{m}=(0.0083\pm0.0068)$~c/keV/kg/d. The COSINE-100 counting rate in the 1-6~keV$_{ee}$ region for the crystals considered has been evaluated as 2.7~c/keV/kg/d~\cite{cosinebkg2021}. COSINE-100 is running and a 3$\sigma$ coverage of the DAMA/LIBRA region was foreseen for five years of data with full exposure. In parallel, in-house crystal manufacturing in order to reduce intrinsic radioactivity is ongoing in Korea for a new phase of the experiment using  $\sim$200~kg of NaI(Tl) (COSINE-200) \cite{cosinefacility}; first results for small crystals (with a mass of $\sim$0.7~kg) have shown a decrease of $^{210}$Pb activity  \cite{cosinecrystals}.

\begin{figure}[]
\centering
\includegraphics[width=12 cm]{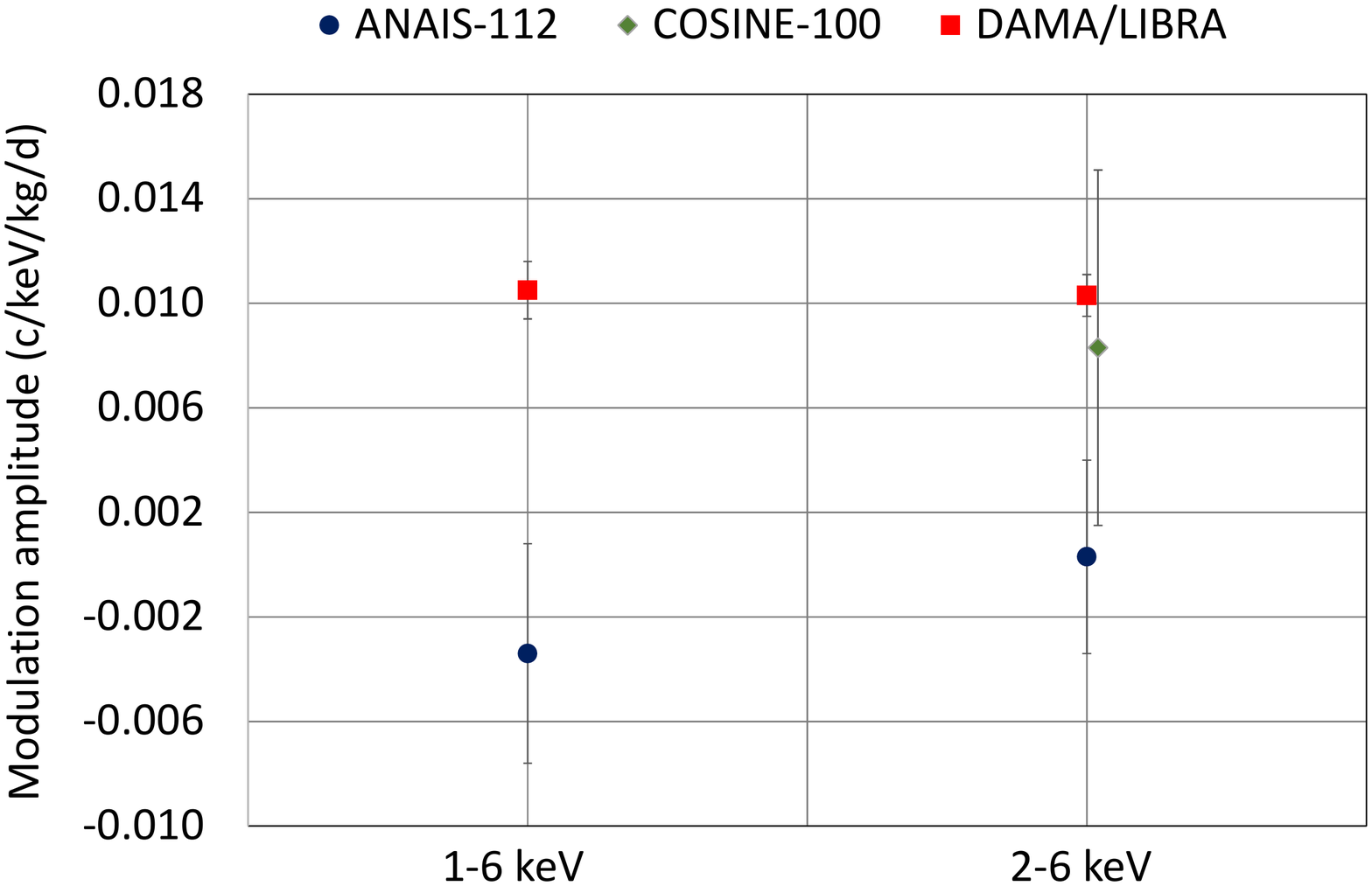}
\caption{Comparison of the modulation amplitudes $S_{m}$ derived  by DAMA/LIBRA \cite{damaphase2}, ANAIS-112~\cite{anais3y} and COSINE-100 \cite{cosinemod} experiments for the 1-6~keV$_{ee}$ and 2-6~keV$_{ee}$ energy regions.}
\label{plotam}
\end{figure}

The comparison between the modulation amplitudes obtained by DAMA/LIBRA, ANAIS-112 and COSINE-100 experiments is shown in Fig.~\ref{plotam}.

The SABRE project (Sodium-iodide with Active Background REjection) is being prepared at LNGS \cite{sabre}; twin detectors in the northern and southern hemispheres (at LNGS and at the Stawell Laboratory, in construction in Australia, respectively) will be set-up in the end to analyze any possible seasonal effects induced by backgrounds, which would have opposite phase in those detectors. SABRE is working on the procurement of NaI(Tl) crystals with improved radiopurity; a potassium concentration of (4.3$\pm$0.2)~ppb has been measured by ICPMS for a crystal \cite{sabreK} and further developments are in progress \cite{nai33}. A target mass of $\sim$50~kg is planned to be operated using not only passive but also active shielding (with a veto made of liquid scintillator). A reduction of the background rate obtained by DAMA/LIBRA in the region of interest of about one order of magnitude is intended; Monte Carlo simulations for expected background have been made \cite{ANTONELLO20191}. For an exposure of three years, DM-nucleon cross sections up to to $2\times10^{-42}$~cm$^{2}$ could be explored for masses around $\sim$45~GeV/c$^{2}$. Tests with a 3.5~kg crystal shipped from US to LNGS are ongoing (SABRE Proof of Principle, PoP).

The PICOLON (Pure Inorganic Crystal Observatory for LOw-energy Neutr(al)ino) project, placed at Kamioka, is also working on the development of NaI(Tl) detectors with ultra-high purity following recristallization and using appropriate resins, in order to reduce particularly $^{210}$Pb and $^{40}$K, and satisfactory tests with small crystals have been made \cite{picolon,picolon2}. A first phase using four modules of NaI(Tl) with a total mass of 23.4~kg could be prepared in 2021 \cite{picolon2}. In a longer term and after completing various steps, the installation inside the KamLAND detector of a large mass (hundreds of kg) of NaI(Tl) is being considered. \cite{KOZLOV2020162239}.

NaI(Tl) crystals were firstly considered as bolometers, operated at low temperatures, in \cite{coron2013}. The COSINUS experiment (Cryogenic Observatory for SIgnatures seen in Next-generation Underground Searches), located also at LNGS, is based on a different detection approach than the other NaI(Tl) projects, by developing scintillating bolometers made of NaI, following the CRESST technology~\cite{cosinus}. The phonon signal does not depend on the type of particle generating the event but the scintillation light does, so a detector measuring simultaneously heat and light can discriminate NR and ER on an event-by-event basis. This capability has been shown with crystals having masses of tens of grams. Pure NaI could be used, thanks to the intrinsic scintillation at cryogenic temperatures. If COSINUS excluded a DM interaction rate of $\sim$0.01~cpd/kg above 1.8~keV, the DM interpretation of DAMA/LIBRA signal would be ruled out~\cite{cosinusjcap}. The construction of the experiment is underway and the first results on DM are expected for 2023.

Summarizing, several projects using sodium iodide as DM target are working to clarify the long-standing puzzle of the annual modulation observed {over two decades} by DAMA/LIBRA, being in conflict with {exclusion results assuming certain scenarios} from other experiments using different targets. ANAIS-112 and COSINE-100 have already presented very relevant results, although still with limited significance; {the derived modulation amplitude is compatible with zero from 3 years of data of ANAIS-112, although from 1.7 years of data of COSINE-100 it is also compatible with that of DAMA/LIBRA. Other} projects with particular, attractive features are also underway. The main features of all of them are summarized in Table~\ref{tablemod}.

\begin{table}
\caption{Summary of the main features of sodium iodide experiments and projects focused on the DM annual modulation signature.}
\label{tablemod}
\centering
\begin{tabular}{lcccccc}
\toprule
\textbf{Experiment}	& \textbf{Laboratory}	& \textbf{Technology} 	&  \textbf{Target} & \textbf{Size} 	& \textbf{Status} 	& \textbf{Reference} \\
\midrule
DAMA/LIBRA & LNGS  &  Scintillator & NaI(Tl) & $\sim$250~kg & Running  & \cite{damaphase2,bernabeippnp2020}\\
ANAIS-112 & LSC & Scintillator & NaI(Tl) & 112.5~ kg & Running & \cite{anaismod,anaistaup} \\
COSINE-100 & Yangyang & Scintillator & NaI(Tl) & 106~kg & Running & \cite{cosinemod}\\
SABRE & LNGS,Stawell  &  Scintillator & NaI(Tl) & $\sim$50~kg & In preparation  & \cite{sabre} \\
PICOLON & Kamioka & Scintillator & NaI(Tl) & 23.4~kg & In preparation  & \cite{picolon} \\
COSINUS & LNGS & Bolometer & NaI,NaI(Tl) & $\sim$1~kg & In preparation  & \cite{cosinus} \\
\bottomrule
\end{tabular}
\end{table}

\section{Signal directionality}
\label{secdir}

As the Sun turns around the center of our galaxy, the wind of DM particles moving through the solar system is coming mainly from the Cygnus constellation. Then, the recording of the direction of tracks produced by NR could allow to demonstrate the galactic origin of a potential signal by discriminating DM events from backgrounds, expected to produce isotropic tracks \cite{Spergel:1987kx,Mayet:2016zxu,ohare2021}.
It is estimated that the difference for the forward-backward rates from DM interactions could be about a factor 10 (depending on the energy threshold). Moreover, after a detection of a DM signal, many properties of DM like their velocity distribution could be explored in this way, opening the possibility of the so-called DM astronomy. Directional DM experiments could explore cross sections below the limits set by neutrino interactions (neutrino floor), even having smaller exposures than other types of experiments, thanks to the different angular distributions of the recoils induced by DM or by neutrinos; atmospheric and supernova neutrinos would give roughly isotropic recoils, while those from solar neutrinos would concentrate in the opposite direction from the Sun. Moreover, directional detectors could be used not only for the direct detection of galactic DM particles but also to search for light DM behind an electron accelerator beam dump \cite{PhysRevD.99.061301} and to detect supernova-produced DM (new particles with mass on the MeV scale with a flux steeply peaked towards the galactic center) \cite{PhysRevD.102.075036}.

The reconstruction of tracks in detectors is really challenging, since for NR with energies at the keV scale, paths are very short: $\sim$1~mm in gas media and $\sim$0.1~$\mu$m in solid material. To fully determine the direction, it would be necessary to register axis and sense; but the observation of the so-called head-tail asymmetry, by registering the relative energy loss along the path, can be helpful. In addition, a daily modulation on the incoming direction of DM particles, due to the Earth's rotation, is also expected. 

Two approaches are being followed in the construction of directional detectors for DM. On one hand, using  nuclear emulsions, and on the other, operating gas targets at low pressure ($\sim$0.1~atm) inside TPCs, having different charge amplification systems and track readout devices, like MWPCs (Multi-Wire Proportional Chambers), MPGDs (Micro Pattern Gaseous Detectors) and optical readouts \cite{battat}. Low pressure gas targets are required to have longer tracks which could be more easily reconstructed. Then, the mass of the DM target accumulated in these experiments is typically much lower than in other DM detectors. Due to a different range and ionisation density, the discrimination between NR and ER is in some cases possible. It is worth noting that in most of the gas detectors, mixtures containing $^{19}$F, with non-null spin, are used. In this section, the main projects working on all these approaches will be presented; Table \ref{tabledir} shows a summary of their main properties.

\begin{table}
\caption{Summary of the main features of projects focused on the directional DM detection.}
\label{tabledir}
\centering
\begin{tabular}{lcccccc}
\toprule
\textbf{Experiment}	& \textbf{Laboratory}	& \textbf{Technology} 	&  \textbf{Target} & \textbf{Size} 	& \textbf{Status} 	& \textbf{Reference} \\
\midrule
DRIFT &  Boulky  & TPC+MWPC  & CS$_{2}$+CF$_{4}$+O$_{2}$ & 0.14~kg, 1~m$^3$  & Finished & \cite{battat3}\\
MIMAC & LSM  & TPC+Micromegas & CHF$_{3}$+CF$_{4}$+C$_{4}$H$_{10}$ & 1~m$^3$ & In preparation & \cite{riffard,TAO2021164569} \\
NEWAGE & Kamioka  & TPC+$\mu$PIC &  CF$_{4}$, SF$_{6}$ & 0.01~kg & Running & \cite{nakamura,yakabe}\\
DMTPC & WIPP  & TPC+opt. read. & CF$_4$ & 1~m$^3$ & In preparation & \cite{deaconu} \\
CYGNO & LNGS  & TPC+GEM,CMOS,PMT & He/CF$_4$ & 1~m$^3$ & In preparation & \cite{cygno,costa} \\ ine
NEWS-dm & LNGS & Nuc. emulsion+opt. read  & Silver halide & 10~g & In preparation & \cite{newsdm} \\
\bottomrule
\end{tabular}
\end{table}

\subsection{Time Projection Chambers}

The DRIFT experiment (Directional Recoil Identification From Tracks) can be considered as the pioneer of directional DM detectors. It consisted of a TPC, having a conversion volume of around 1~m$^{3}$, equipped with two attached MWPCs sharing a central cathode. The stainless steel vessel was filled with an electronegative gas, using a mixture made of CS$_{2}$+CF$_{4}$+O$_{2}$ (corresponding to 0.14~kg at a pressure of 55~mbar); then, the produced ions, instead of the electrons, are drifted towards the readout, which helps to reduce the effects of diffusion and to optimize the resolution of tracks. This approach is scalable to large volumes, although the spatial granularity is limited (larger than $\sim$1~mm). Background from ER can be rejected, but that from alpha particles is problematic. DRIFT operated at the Boulby Underground Laboratory, in UK, for more than ten years. Directional NR, induced by neutrons from a $^{252}$Cf source, were measured and the head-tail asymmetry was quantified \cite{battat2}. Constraints for SD DM-proton interaction were derived from an exposure of 54.7~days, down to 2.8$\times$10$^{-37}$~cm$^2$ for a mass of 100~GeV/c$^2$ \cite{battat3}{, which have recently been improved in the low mass region thanks to machine learning algorithms to better discriminate signal and background events} \cite{drift2021}. 

The MIMAC experiment (MIcro-tpc MAtrix of Chambers) also uses a dual TPC with a shared cathode, which in this case is equipped with pixelized Micromegas (Micromesh Gas structures). The bulk Micromegas technology used offers stability and good spatial and energy resolution. MIMAC operates at the LSM since 2012. The stainless steel vessel is filled with a CHF$_{3}$+CF$_{4}$+C$_{4}$H$_{10}$ mixture, working at $\sim$50~mbar. A low threshold of 2~keV$_{ee}$ has been achieved in a prototype. Quenching factors for F and He have been measured using directly an ion source at Grenoble. 3D tracks of NR from radon were also registered \cite{riffard} and the observation with angular resolution lower than 10$^{o}$ of $^{19}$F ion tracks (from an ion beam) has been achieved \cite{tao,TAO2021164569}. A detector with volume at the scale of 1~m$^{3}$ is in preparation.

The NEWAGE experiment (NEw generation WIMP search with an Advanced Gaseous tracker Experiment) is based on a monolithic system with a TPC equipped with a micro-pixel ($\mu$-PIC) chamber for amplification and readout. After operating on surface, long data-taking at the Kamioka underground laboratory was made filling the vessel with CF$_{4}$ gas at $\sim$100~mbar. NEWAGE also demonstrated the 3D detection of NR tracks, observed the head-tail effect at energies larger than 100~keV and presented first results on DM interactions \cite{nakamura}. Limits for the SD DM-proton interaction cross section have been derived with data taken from 2013 to 2017 corresponding to 4.51~kg$\cdot$d, achieving a value of 4.3$\times$10$^2$~pb for WIMPs with mass of 150~GeV/c$^2$ \cite{yakabe,newage2021}. Since 2018, a detector with low background specifications using polyimide is running \cite{HASHIMOTO2020164285}. Developments with SF$_{6}$ gas in a negative ion micro-TPC detector are also underway~\cite{sf6}.

The DMTPC experiment (Dark Matter Time-Projection Chamber) uses a TPC with both charge and optical (PMTs, CCDs) readouts providing complementary energy measurement; CCDs offer high granularity and quite easy data acquisition. Since 2007, various prototypes of 10 and 20~liters have been operated, firstly at MIT and later underground at WIPP (Waste Isolation Pilot Plant) facilities, in US. The vessel was filled with CF$_4$ gas working at low pressure (30-100 Torr). The direction of NRs was measured in the 20~liter chamber \cite{deaconu}. Now, a detector at the 1~m$^{3}$ scale is in preparation, using four TPCs.

CYGNUS is a proto-collaboration which has been set-up gathering many of the groups focused on directional DM detection; the aim is to develop a multitarget and multisite observatory of DM with capabilities to measure directionality \cite{cygnus}. Particular common goals are the study of different mixtures of gas, the reduction of the energy threshold below 1~keV$_{ee}$ in these type of detectors and the increase of the detector volumes (from 10 to even 1000~m$^3$). Both high and low density gas targets can be considered, to work respectively either on search mode or with directionality. The science case for such a large NR observatory, a description of technologies and careful estimates of the expected sensitivity under different scenarios are presented in detail at \cite{cygnus}. There are projects underway to install CYGNUS detectors at several laboratories in US, UK, Japan, Italy and Australia and first prototypes are already being operated. For instance, CYGNO \cite{cygno}, at LNGS, has developed some small detectors using He/CF$_{4}$ and read by PMTs, CMOS cameras and GEMs \cite{costa,cygno2}; NRs, produced by neutrons, registering direction and sense have been observed in the LEMOn (Large Elliptical Module Optically readout) prototype \cite{lemon} and the achievement of a 2~keV$_{ee}$ energy threshold has been shown. The immediate goal is to build a 1~m$^{3}$ apparatus, CYGNO/INITIUM, using the same technologies and with a target mass of $\sim$1~kg.

\subsection{Other techniques}

The NEWS-dm project (Nuclear Emulsions for WIMP Search with directional measurement) follows a different detection approach \cite{newsdm}. The target for DM and the tracking detector are made of an emulsion film having silver halide crystals immersed in a polymer; then, NRs generate silver clusters (at the nm scale) and the 3D tracks can be reconstructed using an optical microscope. NIT (Nano Imaging Tracker) emulsions using nanometric grains have been developed and new scanning systems, fully automated, to overcome diffraction limitations are being designed. The achievement of an excellent spatial resolution of 10~nm has been shown. Tests with a 10~g prototype are underway at LNGS in order to quantify backgrounds. The preservation of the direction of DM particles by NR has been studied by simulation using the SRIM code considering different recoil atom types finding a very good performance \cite{emulsion2021}. In a longer term, the operation of a detector put on an equatorial telescope (absorbing the rotation of the Earth) to follow the Cygnus constellation for 10 kg$\cdot$year has been proposed.

Together with all these efforts on directional detectors, other proposals have been made or are in development to determine the recoil direction in DM detectors, based on very different approaches:
\begin{itemize}
    \item Following crystal defect spectroscopy, DM-induced NRs in a target made of diamond would produce an observable damage trail in the crystal altering the strain pattern \cite{rajendran,diamond2}. An ultra-fine spatial resolution at the nm-scale is expected and, being a solid-state detector, a large target mass could be accumulated.
    \item A DNA strand detector could be implemented \cite{dna}; DNA strands prepared onto a nm-thick gold foil would be severed by the recoil of a gold atom, kicked out by a DM particle. The identification of the location of each severing event could be determined applying biological techniques.
    \item Planar targets like graphene make easier the determination of recoil directions as multiple interactions in the bulk target are avoided \cite{graphene}. In PTOLEMY  \cite{ptolemy}, graphene FETs in stacked planar arrays with tunable meV band gaps would offer single-electron sensitivity.
    \item In paleo-detectors, traces left in ancient minerals by the DM interaction could be searched for, taking advantage of huge integration times \cite{PhysRevD.99.043014}. Different readout scenarios are being considered to achieve nm resolution and the mineral selection can be optimized to suppress some cosmogenic and radiogenic backgrounds. 
    \item The dependence of the columnar recombination on the alignment of the NR direction with respect to a drift field provides a directional sensitivity, being investigated in the ReD (Recoil Directionality) project within the DarkSide Collaboration. This has been explored for a double-phase argon TPC \cite{Cadeddu:2017ebu}. 
    \item More ideas are being developed considering for instance carbon nanotubes \cite{cnt} or anisotropic crystals like GaAs and sapphire \cite{PhysRevD.98.115034} or scintillators like ZnWO$_4$.
\end{itemize}

In summary, different projects for directional DM detection are underway all over the world exploiting different techniques. Due to the technological challenges these experiments must face, sensitivities worse than in other types of detector have been reached until now for exploring DM interaction cross sections {and less stringent exclusion curves have been set for SI or SD couplings.} But medium-size prototypes up to 1~m$^{3}$ have been built or are in preparation, achieving important progress on fundamental requirements such as radiopurity, scalability or stability.

\section{Low-mass DM}
\label{seclm}

Experiments intended to explore specifically potential light DM particles (below the GeV/c$^2$ scale) have additional requirements: lighter targets must be used to guarantee the kinematic matching between the DM particles and nuclei; lower energy thresholds are needed as even smaller signals are expected; and other new search channels have to be considered. As a light DM particle cannot transfer enough momentum to produce a detectable NR, scattering off not only by nuclei but also by electrons and absorption are being taken into account. It is worth noting that for low mass DM, the sensitivity is enhanced down to the MeV/c$^2$ scale when considering the Migdal effect \cite{Ibe:2017yqa}; it has been proposed that the DM-nucleus interaction could produce ionization or excitation of recoiling atoms (as electrons do not follow instantaneously the NR), which would give an additional ER signal particularly relevant for DM particles with low mass. The detection of bremsstrahlung emissions accompanying an undetectable NR is being also considered. {It must be noted that the existence of these effects has still to be experimentally proven.}

This need of extremely low energy thresholds and new detection channels has compelled over the last years the proposal and development of ultra-sensitive detection ideas based on advanced technologies to explore different mass ranges \cite{whitepaper}. Experiments using ionization detectors, based on semiconductor devices or noble gas chambers, are producing very interesting results and will be mainly described here. Table \ref{tablelm} presents a summary of their main properties.

\begin{table}
\caption{Summary of the main features of experiments and projects focused on low mass DM using purely ionization detectors.}
\label{tablelm}
\centering
\begin{tabular}{lcccccc}
\toprule
\textbf{Experiment}	& \textbf{Laboratory}	& \textbf{Technology} 	&  \textbf{Target} & \textbf{Size} 	&   \textbf{Status} 	& \textbf{Reference} \\
\midrule
CDEX-10 & Jinping & Point-Contact Ge & Ge & $\sim$10~kg  & Running & \cite{Liu:2019kzq,Jiang:2018pic} \\
DAMIC & SNOLAB & CCD & Si & $\sim$40~g & Running & \cite{damic3} \\ 
DAMIC-M & LSM & Skipper CCD & Si & $\sim$700~ g & In preparation & \cite{damicm} \\
SENSEI & SNOLAB & Skipper CCD & Si & 100~g & In preparation & \cite{sensei,PhysRevLett.125.171802} \\ ine
SEDINE & LSM & Spherical Proportional Counter & Ne-CH$_{4}$ & $\sim$300~g & Finished & \cite{newsg1} \\ 
NEWS-G &  SNOLAB & Spherical Proportional Counter & H,He,Ne & $\sim$1.4~m$^3$ & In preparation & \cite{giroux} \\
TREX-DM & LSC & TPC+Micromegas & Ar/Ne & 300/160~g & In preparation & \cite{trexdmiguaz,trexdmbkg} \\
\bottomrule
\end{tabular}
\end{table}

\subsection{Semiconductor detectors}

As demonstrated by the CoGeNT experiment with a 440~g detector operated at the Soudan Underground Laboratory \cite{Aalseth:2012if}, Point-Contact germanium detectors can achieve sub-keV energy thresholds (due to the small capacitance) combined with excellent intrinsic radiopurity and a massive DM target, although the discrimination between ER and NR is nor possible. The CDEX experiment (China Dark matter EXperiment) is following this approach at the Jinping underground laboratory. CDEX-1 used two detectors with a mass of $\sim$1~kg each, with a threshold of 160~eV$_{ee}$. An annual modulation analysis (from 3.2~years of data and an energy threshold of 250~eV$_{ee}$ threshold) was made finding no signal \cite{cdexmod} and limits on DM-nucleon SI and SD couplings, and considering the Migdal effect, were derived \cite{Liu:2019kzq}. An array of 3$\times$3 detectors immersed in liquid N$_{2}$ with a total mass of $\sim$10~kg is used in CDEX-10; results for both SI and SD interactions from 102.8~kg$\cdot$d data and an analysis threshold of 160~eV$_{ee}$ have been already released \cite{Jiang:2018pic} as well as constraints on dark photons \cite{PhysRevLett.124.111301} and on WIMP couplings within the framework of effective field theories \cite{cdexeft}. The preparation of larger set-ups (CDEX-100 and CDEX-1T) is underway with home-made germanium  detectors to be placed inside a 1700~m$^3$ liquid N$_{2}$ tank. Point-Contact germanium detectors (n-type) are also used in the TEXONO project, which has set constraints on bosonic DM from 314.25~kg$\cdot$d and upper limits on electron couplings \cite{SINGH201963}.

In Charge-Coupled Devices (CCDs) made of silicon, the charge generated by the interaction drifts towards the pixel gates until being readout. As the interaction depth is correlated with the lateral charge diffusion, in this approach the 3D position can be reconstructed and effective particle identification is possible for background discrimination. The DAMIC experiment (DArk Matter In CCDs) operates 7 high resistivity, fully depleted CCDs ($\sim$6~g each) at SNOLAB since 2017; a leakage current at 2~e$^{-}$/mm$^{2}$/day has been achieved, with an energy threshold of 50~eV$_{ee}$. A precise determination of the quenching factor in Si down to 60~eV$_{ee}$ (using a Sb-Be source generating 24~keV neutrons) was made and the radioactive impurities in silicon bulk were analyzed through time-correlated spatial coincidences \cite{matalon}. Limits for low mass DM considering interactions on electrons and also hidden photon DM were obtained \cite{Aguilar-Arevalo:2016zop,PhysRevLett.123.181802} and updated results on nucleon scattering from 11~kg$\cdot$d have been presented \cite{damic3}; above the threshold, an excess of charge events has been observed, which still requires  investigation. DAMIC-M \cite{damicm} will be operated at LSM using $\sim$50 more massive CCDs (with 13.5~g each) based on the Skipper readout, where through the multiple measurement of the pixel charges, noise is reduced and single electron counting at high resolution is achieved, as it has already been shown. It will be able to observe collision energies of only 1~eV. Commissioning of DAMIC-M in Modane is expected at 2023 and the goal is to achieve a background of 0.1~c/keV/kg/d, thanks to the control of cosmogenic $^3$H and surface $^{210}$Pb. The cosmogenic activation of silicon has been carefully analyzed \cite{PhysRevD.102.102006}. For a 1~kg$\cdot$y exposure, DAMIC-M will explore SI DM-nucleon scattering at the GeV/c$^2$ range, DM-electron interactions from 1~MeV/c$^2$ to 1~GeV/c$^2$ and DM candidates from the hidden sector \cite{damicm}.

The novel Skipper readout is being already used by the SENSEI experiment (Sub-Electron-Noise Skipper CCD Experimental Instrument); a new generation of CCDs consisting of million pixels has been developed and small prototypes (with total active masses of 0.0947 and 2~g) have been operated at Fermilab MINOS Hall in US (placed 100~m underground). SENSEI has set the most restrictive constraints on DM-electron scattering below a few MeV/c$^2$ \cite{sensei,PhysRevLett.125.171802} and a broad range of hidden-sector DM candidates can also be explored. The installation deep underground at SNOLAB of a 100-g detector (using 48~CCDs) with custom-designed electronics has been proposed. In a longer term, the Oscura project plans to operate 10~kg of Skipper CCDs.

\subsection{Gas detectors}

A Spherical Proportional Counter (SPC) is an unconventional gas detector capable of reaching a very low energy threshold due to a reduced capacitance ($<$1~pF) having a large volume~\cite{giomataris}; a small ball at the sphere center acts as the anode, surrounded by the avalanche region. The SEDINE detector, operated at LSM, consisted of a sphere (60~cm diameter) made of NOSV copper and filled with a gas mixture of Ne-CH$_4$(0.7\%) at a pressure of 3.1~bar (giving an active mass of $\sim$300~g). An acquisition energy threshold of 50~eV$_{ee}$ was achieved and exclusion results for SI DM interaction were derived from a set of data of 42~day \cite{newsg1}. The NEWS-G experiment (New Experiments With Spheres-Gas) will operate at SNOLab a larger sphere (140~cm diameter) built in France and tested at LSM \cite{giroux}. A thorough work has been made to mitigate $^{210}$Pb contamination in the copper sphere by electroplating a thin layer of ultra-radiopure copper onto the inner surface of the detector \cite{BALOGH2021164844}. Determination of the quenching factors is underway for H from measurements at Grenoble and also at TUNL (Triangle Universities Nuclear Laboratory, Duke University) for Ne-CH$_4$(2\%) down to 300~eV$_{ee}$. The response of the detector to single electrons (drift/diffusion times, gain, \dots) was studied using a laser system~\cite{newsg2} and tests of different sensors are ongoing \cite{Katsioulas:2018pyh,Giomataris:2020rna}. Several light targets (H, He) are being considered and commissioning data with CH$_4$ at 135~mbar (mass of 110~g) were taken at LSM using the new large sphere before transportation to SNOLAB, where commissioning is ongoing. Thanks to the  background reduction, better sensor performance and also improved analysis methods \cite{giroux,Katsioulas:2020ycw}, and considering  Ne-CH$_4$(10\%), DM-nucleon cross sections down to 10$^{-41}$~cm$^2$ are expected to be explored for a mass of the DM particle of 1-2~GeV/c$^2$. The development of electroformed copper spheres is being studied in collaboration with PNNL (Pacific Northwest National Laboratory); the design of ECUME, a 140~cm in diameter fully electroformed underground SPC, is underway and construction in SNOLAB is foreseen for 2021. In a longer term, the construction of a larger sphere (3~m diameter), DarkSPHERE, is being investigated; using He-C$_4$H$_{10}$(10\%), the projected sensitivity reaches the neutrino floor in the sub-GeV mass range.

Other types of gas detectors are being developed too. The TREX-DM experiment (Tpcs for Rare Event eXperiments-Dark Matter) uses a gas TPC equipped with Micromegas readouts \cite{trexdmiguaz,trexdmigor,trexdmbkg}, offering low intrinsic radioactivity and the possibility of recording topological information to discriminate backgrounds. The pressurized gas at 10~bar is held inside a vessel (with diameter and height of 0.5~m) made of 6-cm-thick copper and the largest microbulk Micromegas planes ever built (25$\times$25~cm$^2$) are being used. The detector is being presently commissioned at LSC. Being the target flexible, runs using Argon and Neon mixtures with isobutane have been performed below 8~bar; projections for the threshold range from 0.4~keV$_{ee}$ to 0.1~keV$_{ee}$. A complete background model has been made, predicting levels of a few c/keV/kg/d in the region of interest \cite{trexdmbkg}.

\subsection{Other techniques}

The stringent requirements to investigate low mass DM have pushed the development of novel technologies (proposed or already at R\&D phase) aimed at the recording of increasingly smaller energy deposits, the background suppression and the use of light targets: 
\begin{itemize}
    \item In the HeRALD (Helium Roton Apparatus for Light Dark Matter) project, superfluid $^{4}$He has been proposed as DM target \cite{PhysRevD.100.092007}. Sensors consisting of TES (low temperature calorimeters) allow to measure quasiparticles and photons by quantum evaporation (liberation of $^{4}$He atoms into a vacuum). Very low thresholds could be achieved as only 1~meV is necessary to evaporate an He atom. In addition, He offers a light nuclear mass and copious production of scintillation light.
    \item In the SnowBall project, supercooled water (cooled below its normal freezing point) is proposed as DM target, offering the lightest target (H) and the easy availability of water \cite{snowball}. An interacting particle would trigger the water crystallization and a camera is used for the acquisition of image. Tests with neutrons using a 20~g prototype have been made operating at a temperature of $-$20$^{o}$, being insensitive to ER.
    \item Crystals made of laboratory-grown diamond acting as DM target could be outfitted with charge and phonon readouts to resgister the DM scattering (considering both NR and ER) of candidates having very low masses \cite{PhysRevD.99.123005}. Carbon is lighter than other semiconductor materials and a sub-eV theshold can be expected thanks to low noise levels.
    \item Different small band gap materials (at 10-100 meV) from material informatics are being explored as sensors; the use of Si devices with Depleted P-channel Field Effect Transistor (DEPFET) could allow to reach sub-electron noise level to explore MeV DM particles \cite{depfet}. Dirac materials with a small band gap of O(meV) could allow to explore even sub-MeV DM \cite{PhysRevD.97.015004}.
\end{itemize}

Shortly, purely ionization detectors are specially suited for the direct detection of low mass DM. The development of novel technologies for ionization detectors and sensors is ongoing but very low energy thresholds even below 0.1~keV$_{ee}$ have been already achieved in several experiments profiting from very low ionization energies or low capacitance. Although accumulating a large target  mass is not easy in these type of detectors, nuclei with low mass number can be used, like Si in CCDs and He, Ne or Ar in gas detectors. Then, relevant limits on DM-interaction cross sections have been set by these experiments, for nucleon scattering at masses below a few GeV/c$^2$ and, particularly, for electron scattering. In addition, other technologies are being explored in the path to detect the tiny energy deposits expected from very low mass DM particles.

\section{SD interactions}
\label{secsd}

Bubble chambers operate using as targets metastable superheated liquids; then, following a local phase transition, dense energy depositions create bubbles which can be counted and localized with precision at mm by cameras. They are threshold detectors, as the recoil energy cannot be directly determined; dead times are large due to the required time to compress and decompress the detector after an interaction. However, these detectors are virtually unaffected by backgrounds producing ER as the bubble formation probability can be tuned so that only NR generate bubbles. Different target fluids can be used in bubble chambers, typically refrigerants; as most of them contain $^{19}$F (for instance CF$_3$I, C$_3$F$_8$ or C$_4$F$_{10}$), they have the highest sensitivity to SD interactions on protons. This type of detectors has been therefore focused on the study on SD DM couplings.

PICO joined together in 2012 the PICASSO and COUPP experiments and is operating since then a series of increasingly larger bubble chambers at SNOLAB. Following the first PICO-2L, in PICO-60, using 52~kg of C$_{3}$F$_{8}$, an energy threshold of 2.45~keV$_{nr}$ was obtained and the present best limits from direct detection experiments have been derived on SD DM-proton interaction cross sections~\cite{Amole:2019fdf}. For PICO-40L, some changes have been implemented in the design, like the buffer-free concept using no water inside the inner vessel to suppress some backgrounds and with the chamber built ``right-side-up''~\cite{picobufferfree}; data taking with this detector is already starting. For the future, PICO-500 will be a tonne-scale detector with sensitivity to explore proton interaction cross section of 10$^{-42}$~cm$^2$ for a mass of the DM particle of tens of GeV/c$^2$.

Developments are underway to operate scintillating bubble chambers, combining the extreme electron rejection and simple instrumentation of a bubble chamber with the event-by-event energy resolution of a liquid scintillator like Argon or Xenon. The technique has been established for a 30~g xenon bubble chamber \cite{scintbubcham}. The SBC (Scintillating Bubble Chamber) project is preparing at Fermilab such a detector using a LAr chamber with 10~kg, which could be operated at SNOLAB in 2023.

\section{Outlook and Summary} \label{secsum}

In the quest of the DM particles which can be populating our galactic halo, one of the approaches followed is to register the scattering of those particles in a suitable detector. For this direct detection of DM, experiments based on noble liquids or solid-state cryogenic detectors, with increasing target mass and sophistication, have produced very relevant (even if negative) results for years. But complementary searches are developed based on other detection technologies; smaller scale experiments, focused on different physics cases, are indeed a very active field too. 

The identification of distinctive signatures of DM interaction, like the annual modulation in the rates or the signal directionality, would help to confirm a DM detection. Relevant results have been presented by NaI(Tl) experiments, ANAIS-112 and COSINE-100, aimed to independently verify using the same target the DAMA/LIBRA annual modulation result observed over two decades; there are underway other NaI(Tl) projects too with attractive particular features. {Although having higher background levels and energy threshold than other detectors, NaI(Tl) scintillators offer the robust and stable operation and the possibility of accumulating large target mass required in the investigation of the DM annual modulation signal.}

The construction of a DM detector sensitive to directionality (and then capable of proving the galactic origin of a potential signal) is hard, as short tracks must be imaged with good resolution. Low-pressure TPCs equipped with different readouts (MWPCs, Micromegas, $\mu$PICs, optical CCDs, \dots) and nuclear emulsions are being implemented. {Modest energy thresholds have been achieved in this type of detectors and collecting large exposures can be difficult, but medium-size TPC prototypes (having volumes from 0.1 to 1~m$^{3}$) have already been} developed and relevant progress is being achieved. 
 
For SI DM-nucleon cross sections, the best limits for masses above a few GeV/c$^2$ come presently from XENON1T; but at lower masses and when considering SD interaction and also electron scattering, the leading results are obtained from several different detection technologies: solid-state cryogenic detectors (scintillating bolometers, Ge and Si crystals), liquid noble detectors (Xe, Ar) operated in S2-only mode (charge collection) but also purely ionization detectors (CCDs, Ge and gas detectors) and bubble chambers. {Although in some cases having a  large target mass is complicated, the} achievement of extremely low energy thresholds, the inclusion of targets with low mass and/or the consideration of several interaction channels are the {assets} for exploring low mass DM; other new technologies are under study to further improve the detector performance. 
 
It has been attempted to show here that small scale experiments have already released very competitive results in the direct detection of DM and, surely, more will come in the next future. Complementary experiments based on different technologies are required to guarantee the exploration of the wide range of possible DM candidate particles.

\funding{This research received no external funding.}

\conflictsofinterest{The authors declare no conflict of interest.}


\externalbibliography{yes}
\bibliography{template}

\end{document}